\documentclass[twocolumn,aps,prl,superscriptaddress,longbibliography]{revtex4-2}
\usepackage[colorlinks,bookmarks=false,citecolor=blue,linkcolor=red,urlcolor=blue]{hyperref}
\usepackage{verbatim}
\usepackage{color}
\usepackage{amsmath,amssymb,bm,braket,float,mathtools}
\usepackage{graphicx,xfrac,appendix}
\usepackage[english]{babel}
\makeatletter

\usepackage{epstopdf}

\makeatother

\begin{document}

\title{Wigner distribution, Wigner entropy, and Anomalous Transport of a Generalized Aubry-Andr\'{e} model}

\author{Feng Lu}
\affiliation{Department of Physics, Zhejiang Normal University, Jinhua 321004, China}

\author{Ao Zhou}
\affiliation{Department of Physics, Zhejiang Normal University, Jinhua 321004, China}

\author{Shujie Cheng}
\thanks{chengsj@zjnu.edu.cn}
\affiliation{Xingzhi College, Zhejiang Normal University, Lanxi 321100, China}
\affiliation{Department of Physics, Zhejiang Normal University, Jinhua 321004, China}

\author{Gao Xianlong}
\thanks{gaoxl@zjnu.edu.cn}
\affiliation{Department of Physics, Zhejiang Normal University, Jinhua 321004, China}

\date{\today}
\begin{abstract}
We investigate generalized Aubry-Andr\'{e} models featuring tunable quasidisordered potentials and a  mobility edge that separates extended and localized states, with critical states for the mobility edge confirmed through finite-size scaling analysis. Numerical results demonstrate that extended, critical, and localized states can be distinguished via their phase-space representations, particularly the Wigner distribution. The associated Wigner entropy, derived from this distribution, peaks at the critical state, enabling precise localization of the mobility edge. Additionally, wave-packet dynamics reveal anomalous transport behaviors, including superdiffusion and subdiffusion, bridging ballistic transport and the absence of diffusion.
%
\end{abstract}

\maketitle

\section{Introduction} 
Anderson localization \cite{Anderson}, a fundamental phenomenon in wave propagation through disordered media, 
continues to be an active field in condensed matter physics. Scaling theory \cite{SPME_1}, applied to disordered systems, 
highlights the critical influence of spatial degrees of freedom on Anderson localization. In one- and two-dimensional systems, 
the introduction of uncorrelated random disturbances results in the exponential localization of all wave functions. 
Consequently, the localization-delocalization transition has been a longstanding focus in low-dimensional disordered systems. 
Conversely, three-dimensional (3D) Anderson system \cite{Anderson,SPME_2,PhysRevLett.134.046302} displays a unique behavior, where wave functions are neither entirely 
localized nor delocalized. An energy threshold, known as mobility edge, separates delocalized states from localized ones. 
However, detecting the mobility edge in 3D systems experimentally remains difficult. Therefore, to gain a deeper understanding 
of the mobility edge, low-dimensional systems—particularly one-dimensional (1D) systems—offer a more feasible research path.

Except for the Anderson systems, the 1D Aubry-Andr\'{e} (AA) model \cite{AA} 
with a self duality
plays a similarly important role in understanding the Anderson localization and the mobility edges as well. The AA model has an extended-localized transition point, which can be exactly derived by the dual transformation. In addition, AA model is also one of the important source for designing systems with mobility edges. 
Biddle and Das Sarma introduced long-range transitions in the AA model \cite{SPME_9,SPME_10}, deriving exact expressions for the mobility edge via a self-dual transformation and verifying them numerically through calculations of the inverse participation ratio. This approach established a paradigm for analytically determining mobility edges in subsequent studies using self-duality \cite{SPME_11,SPME_13,PhysRevLett.131.186303,PhysRevB.108.174202,PhysRevLett.133.226001}, with numerical validation often employing the inverse participation ratio in generalized AA models \cite{PhysRevA.109.043319,SPME_8,PhysRevA.108.033305,Xuzhihao,PhysRevLett.132.236301}. Mobility edges also appear in slowly varying systems, where they can be obtained through semiclassical methods \cite{SPME_6,SPME_7,SPME_8,SPME_13}. More recently, mobility edges have been identified in AA systems with mosaic modulation \cite{PhysRevLett.125.196604}, with exact solutions derived using Avila's global theory \cite{Avila_1,Avila_2}. This analytical framework has since been extended to generalized mosaic models \cite{PhysRevLett.131.176401,PhysRevB.108.064206} and spinful systems \cite{zhou2025_1}. Advances in experimental techniques have enabled the observation of mobility edges via wave-packet dynamics measurements \cite{SPME_exp_1,SPME_exp_2,SPME_exp_3,jiasuotang}.


In the context of wave-packet dynamics, the spreading behavior of wave packets can exhibit five distinct transport regimes: ballistic transport, superdiffusion, diffusion, subdiffusion, and localization \cite{transport_1,transport_2,transport_3,transport_4}. For a generalized AA model featuring a mobility edge separating extended and localized states \cite{Xuzhihao}, prior work has demonstrated that wave packets propagate ballistically in both the extended regime and near the mobility edge, whereas they remain confined without diffusion in the localized regime. In addition to ballistic transport, diffusion, and localization, two anomalous transport behaviors—superdiffusion and subdiffusion—have also been identified \cite{transport_1}. These findings motivate an exploration of whether such anomalous transport phenomena can arise in systems with generalized quasi-disordered potentials and extended-localized mobility edges.
In 1932, Eugene Wigner preposed the phase space representation, i.e., Wigner distribution function 
$W(x, p)$ (here $x$ and $p$ denote the coordinate and momentum, respectively), which enables the concurrent 
description of quantum states or signals in both coordinate and momentum \cite{Wigner_1,Wigner_2,Wigner_3,Wigner_4,Wigner_5,Wigner_6,Wigner_7}. 
It serves as a distinctive “fingerprint” that captures the joint distribution of these variables, hence providing a unique representation for quantum states or signals.
The Wigner distribution function simultaneously encodes information about a quantum state in both position and momentum spaces, a capability not afforded by the inverse participation ratio. Many mathematical methods can be used to convert the experimental data from optical homodyne tomography into the state’s Wigner function \cite{RevModPhys.81.299} , which makes it possible to experimentally measure the Wigner function.
This prompts us to further reveal the difference between the extended state and the localized state in both coordinate and 
momentum dimensions, and to further study how to employ the Wigner distribution function to locate the mobility edge.

The rest of this paper are arranged as follows: In Sec.~\ref{S2} we introduce the generalized AA model with a mobility edge and 
discuss the extended-localized transition. In Sec.~\ref{S3} we introduce the Wigner distribution and discuss the 
correspondence between the Wigner distribution and the wave functions. Besides, we introduce the Wigner entropy and reveal 
that the mobility edge can be located by Wigner entropy. In Sec.~\ref{S4}, we study the quantum transport property of the model. 
Finally, we make a summary in Sec.~\ref{S5}.

\begin{figure}[htp]
  \centering
  \includegraphics[width=0.5\textwidth]{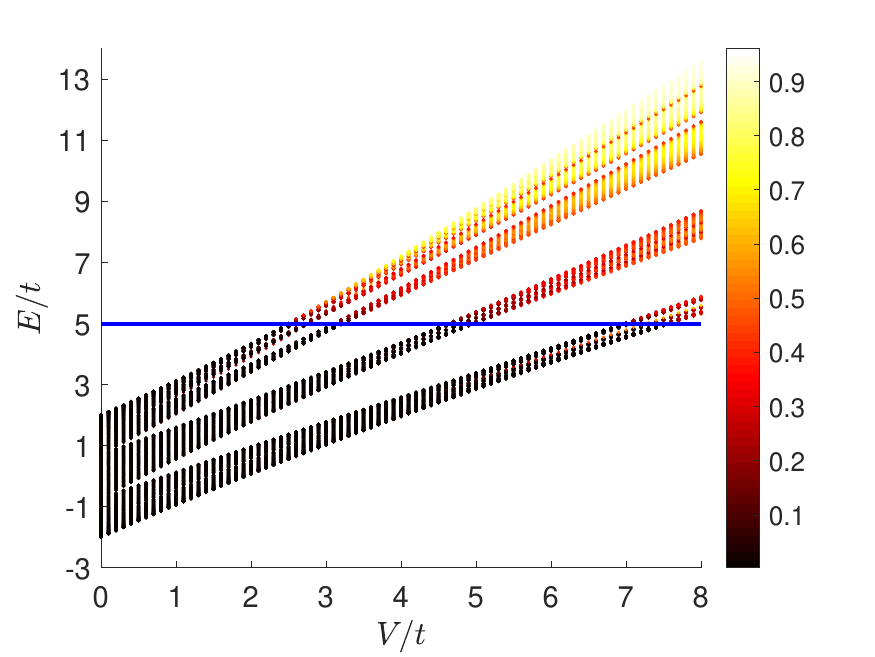}\\
  \caption{(Color Online) The eigenvalues $E$ of Eq.~(\ref{eq1}) versus $V$ under $b=0.4$, $L=500$. 
  The color bar denotes the value of IPR. The blue solid line denotes the invariable mobility edge with $E_c=2t/b=5t$.}
  \label{f1}
\end{figure}

\section{Model and the mobility edge \label{S2}}
We study a 1D generalized AA model with nearest-neighbor hoppings and incommensurate on-site potentials, whose Hamiltonian is 
expressed as 
\begin{equation}
\hat{H}=\sum^{L-1}_{n}t\left(\hat{c}^{\dag}_{n+1}\hat{c}_{n}+\hat{c}^{\dag}_{n}\hat{c}_{n+1}\right)+\sum^{L}_{n}V_{n}\hat{c}^{\dag}_{n}\hat{c}_{n},
\label{eq0}
\end{equation}
where $\hat{c}^{\dag}_{n}$ ($\hat{c}_{n}$) is the creation (annihilation) operator with $n$ the site index, 
$L$ is the length of the system and $t$ is the hopping strength which is set as the unit of energy. 
$V_{n}=\frac{V}{1-b\cos(2\pi\alpha n)}$ is the generally quasidisordered potential, in which $V$ is the strength of the 
potential with $0< b< 1$ being the tuning parameter and $\alpha=\frac{\sqrt{5}-1}{2}$ is the incommensurate parameter 
which is the cause of the quasidisordered potential. 
In coordinate space, we can define the generalized wave function $\ket{\psi}=\sum_{n}\phi_{n}\hat{c}^{\dag}_{n}\ket{0}$ 
with $\phi_{n}$ the amplitude of probability.  Based on $\ket{\psi}$, we can obtain the following  
static Schr\"{o}dinger equation 
\begin{equation}
t\phi_{n+1}+t\phi_{n-1}+\frac{V}{1-b \cos(2\pi\alpha n)} \phi_n=E \phi_n. 
\label{eq1}
\end{equation} 
Employing the self-dual transformation, we can obtain the exact expression of the extended-localized mobility edge $E_{c}$ 
of the generalized AA model \cite{SPME_14}, which satisfying $E_{c}=\frac{2t}{b}$. From the exact expression of $E_{c}$, 
we know that the mobility edge is invariant with $V$, but is uniquely determined by $b$. In the following, we will 
show that in addition to the inverse participation ratio and finite-size scaling analysis to study the extended-localized 
transition over the energy domain, the extended-localized transition can be characterized by the Wigner distribution as well. 
Based on the Wigner distribution function, we can further define the winger entropy and show that the mobility edge can be 
located by winger entropy, and the results are consistent with the analytical solution. Finally, we will show that the system 
with generalized quasidisordered potential has the property of anomalous quantum transport.

\begin{figure}[htp]
  \centering
  \includegraphics[width=0.5\textwidth]{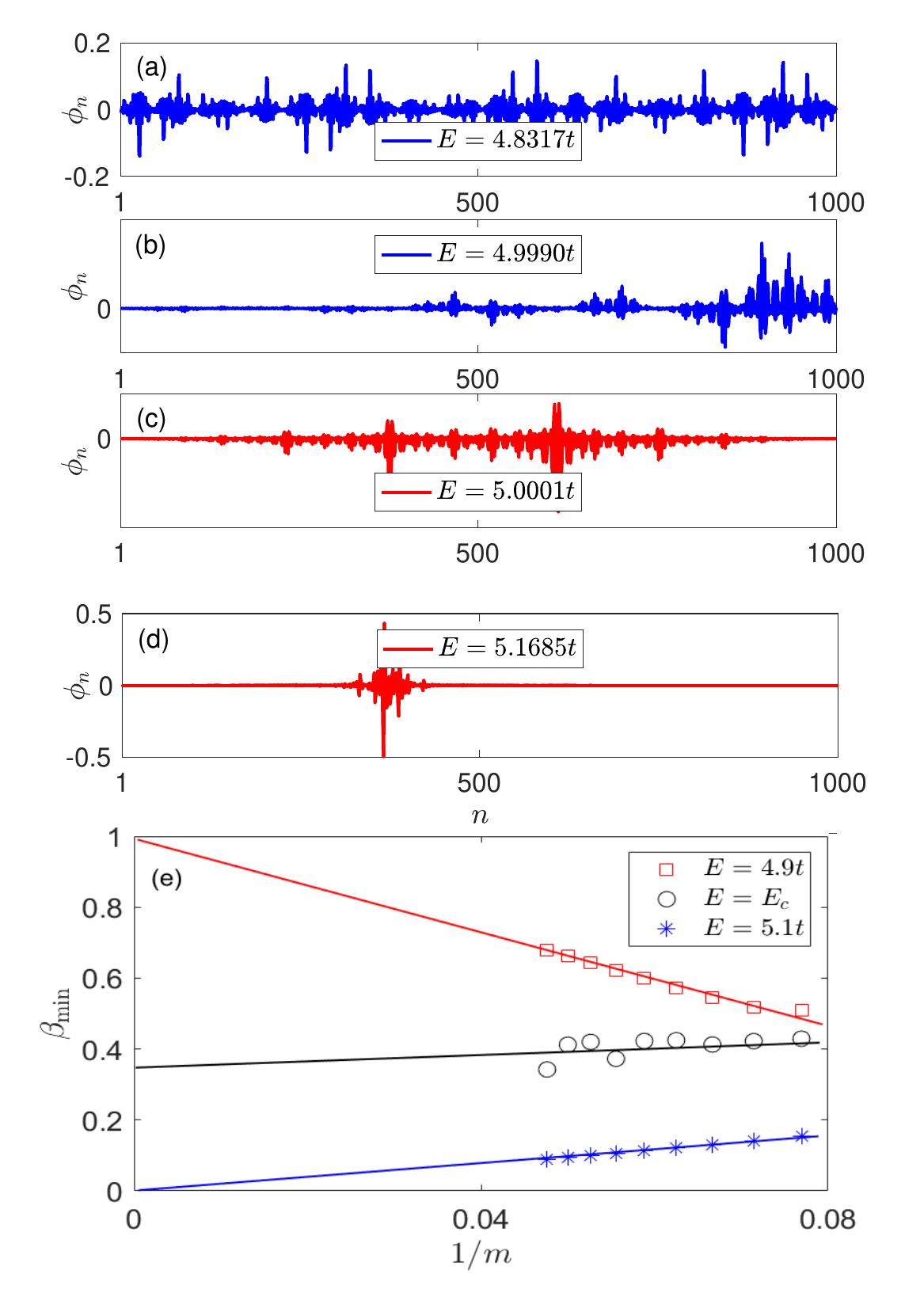}\\
  \caption{(Color Online) (a)-(d) Wave functions obtained from Eq.~(\ref{eq1}) with $b=0.4$, $V=4.8t$ and $L=1000$. 
  Concretely: (a) Extended state far below $E_c$; (b) and (c) critical states at $E_{c}$;  
  (d) localized state far above $E_{c}$. (d) Finite-size scaling analysis on the fractal dimension $\beta_{\rm min}$ at $b=0.4$ 
  and $V=4.8t$. $m$ is the index of the Fibonacci number.
  }
  \label{f2}
\end{figure}

Taking $b=0.4$ and $L=500$, the resulting localization phase diagram is plotted in Fig.~\ref{f1}. The moderate $b$ can present the extended local transformation and mobility edge more intuitively. The 
color bar denotes the value of inverse participation ratio (IPR) defined as ${\rm IPR}_{j}=\sum^{L}_{n=1}|\phi^{j}_{n}|^4$ with 
$j$ the energy level index. For an extended state, ${\rm IPR} \propto L^{-1}$ and decays to zero under larger $L$.  
For a localized state, ${\rm IPR}$ is a finite value \cite{SPME_9,SPME_10}. From Fig.~\ref{f1}, we can see that the invariable 
mobility edge $E_{c}/t=5$ perfectly separates the extended states from the localized state. Below $E_{c}$ 
${\rm IPR}$ tends to $0$, signaling the extended states. Above $E_{c}$, ${\rm IPR}$ are finite values, signaling 
the localized states. The numerical results are consistent with analytical results. To be specific, we plot the typically 
extended and localized state in Figs.~\ref{f2}(a) and \ref{f2}(d), whose energy is far below $E_{c}$ and far above 
$E_{c}$, respectively. We can see that the probability distribution of the extended state extends over all the system 
while that of the localized state only occupies a small part of the system. In addition, at $E_{c}$, the wave functions 
are critical, because we can see from Figs.~\ref{f2}(b) and \ref{f2}(c) that although the probability distribution occupies 
most of system space, there is still a part of the system space that is not occupied, presenting the characteristics of subexpansion. 

We can further verify the preliminary judgment of the spatial distribution of wave function properties at fixed 
system size by finite-size scaling analysis. Here, the finite-size scaling analysis is carried out by calculating 
the fractal dimension $\beta_{\rm min}$. The properties of the wave function are judged by the results of the 
fractal dimension under the extrapolation limit of the large system size. The $\beta_{\rm min}$ can be 
calculated by the fractal theory \cite{fractal_1,fractal_2,fractal_3,fractal_4,fractal_5,fractal_6}. As first, 
we choose a system whose size $L$ is equal to the $m$th Fibonacci number 
$F_{m}$ and the incommensurate parameter $\alpha$ is replaced by the ratio of two neighboring Fibonacci numbers, i.e., 
$\alpha=\frac{F_{m}}{F_{m+1}}$. Then we can extract a scaling index $\beta^{m}_{n}$ from the probability $P^{m}_{n}=|\phi^{m}_{n}|^{2}$ 
by $P^{m}_{n}\sim (1/F_{m})^{\beta^{m}_{n}}$. According to the fractal theory, we know that for an extended state, 
the maximum of $P^{m}_{n}$ follows the scaling ${\rm max}(P^{m}_{n})\sim (1/F_{m})^{1}$, i.e., $\beta_{min}=1$, 
and for a localized state, ${\rm max}(P^{m}_{n}) \sim (1/F_{m})^{0}$, signaling $\beta_{\rm min}=0$, while 
for the critical state, whose $\beta_{\rm min}$ is within the interval (0,~1). Taking $V=4.8t$ and a Fibonacci 
sequence (system sizes), we plot the $\beta_{\rm min}$ of the selected extended (chosen from the lowest eigenstate), 
critical (chosen at $E_{c}$), and localized states (chosen form the highest eigenstate) as a function of $1/m$ in 
Fig.~\ref{f2}(e). We can see that under extrapolation limit ($m\rightarrow \infty$), $\beta_{\rm min}$ 
equals to $1$ for the extended state, and is a finite number within $(0,~1)$ for the critical state, and equals to $0$ 
for the localized state. Therefore, the finite-size analysis validates the preliminary judgement obtained from the 
spatial distribution of the wave functions. 

Having recalled the extended-critical transition and the invariable mobility edge caused by the generalized potentials, in the following section, 
we will focus on employing the Wigner distribution, a phase space pattern, to distinguish extended states and 
the localized states, and employing Wigner entropy to locate the mobility edge. In addition, we will reveal the 
anomalous transport phenomenon of the model. 

\begin{figure}[htp]
  \centering
  \includegraphics[width=0.5\textwidth]{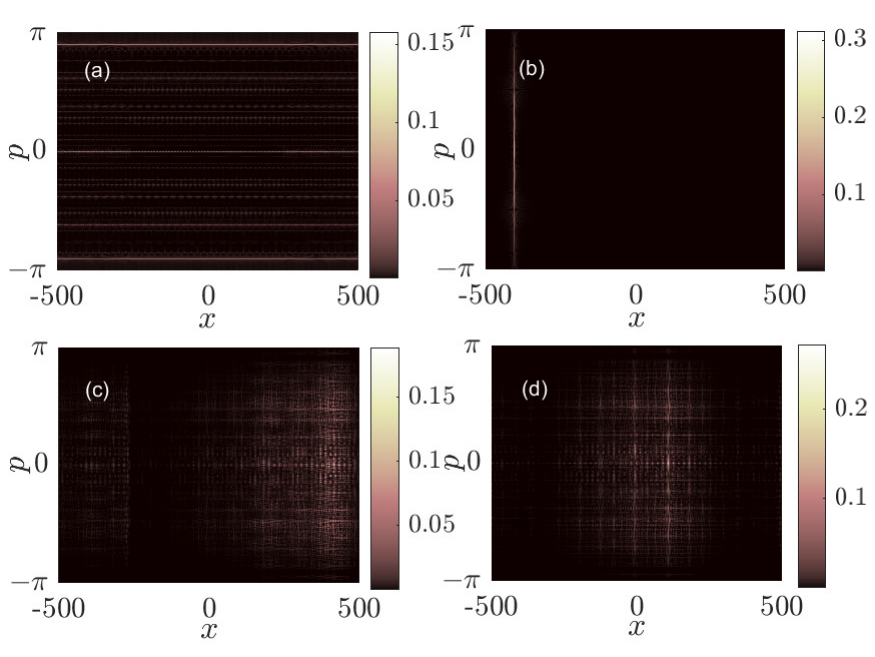}\\
  \caption{(Color Online) The corresponding Wigner distributions of the wave functions under $b=0.4$, $V=4.8t$ and $L=1000$. 
  (a) $W(x,p)$ of the $100$-th wave function; (b) $W(x,p)$ of the $900$-th wave function;  
  (c) $W(x,p)$ of the $520$-th wave function; (d) $W(x,p)$ of the $527$-th wave function. The color bar 
  represents the value of $W(x,p)$. 
  }
  \label{f3}
\end{figure}

\section{Wigner distribution and Wigner entropy}\label{S3}

For the given wave function $\ket{\psi}$, the Wigner distribution function $W(x,p)$ can be 
obtained by the following integration \cite{Wigner_1,Wigner_2,Wigner_3,Wigner_4,Wigner_5,Wigner_6,Wigner_7}
 \begin{equation}
 W(x,p)=\frac{1}{2\pi\hbar}\int^{\infty}_{-\infty}\bra{x-\frac{y}{2}}\hat{\rho}\ket{x+\frac{y}{2}}e^{-\frac{ipy}{\hbar}}dy, 
 \end{equation}
 where $x$ and $p$ are the coordinate and momentum in phase space, respectively, and $\hbar$ is the reduced 
 Planck constant, and $\rho=|\psi\rangle\langle\psi|$. 

The difference between the extended states and localized states can be easily seen from the Wigner distributions. 
We diagonalize the Hamiltonian matrix under $b=0.4$, $V=4.8t$, and $L=1000$. 
The $W(x,p)$ of an extended state (The $100$-th wave function) is plotted in Fig.~\ref{f3}(a). As seen that in the phase space, 
for the extended state, its $W(x,p)$ is extended and continuous in the $x$ branch, while in the $p$ branch, 
the distributions are discrete. For the localized state, the consequence is different. We take the $900$-th wave 
function (localized state) as an example, the corresponding $W(x,p)$ is plotted in Fig.~\ref{f3}(b). In the $x$ branch, the distributions are 
localized, but in the $p$ branch, the distributions are extended and continuous. In fact, for the critical states, 
the results are different from those of the extended and localized states as well. We take the $520$-th and 
$527$-th wave functions (critical states near $E_{c}$) as examples, whose corresponding $W(x,p)$ are plotted 
in Figs.~\ref{f3}(c) and \ref{f3}(d), respectively. We can see that, the distributions are both subextended 
and segmented continuous in the $x$ and $p$ branches. We emphasis that the differences of $W(x,p)$ among extended, 
critical, and localized states are universal. For wave functions of other energy levels, the corresponding Wigner distribution 
results are similar to those in Fig.~\ref{f3}. Therefore, one can effectively employed the Wigner distribution to distinguish 
the extended, critical, and localized states. 

\begin{figure}[htp]
  \centering
  \includegraphics[width=0.5\textwidth]{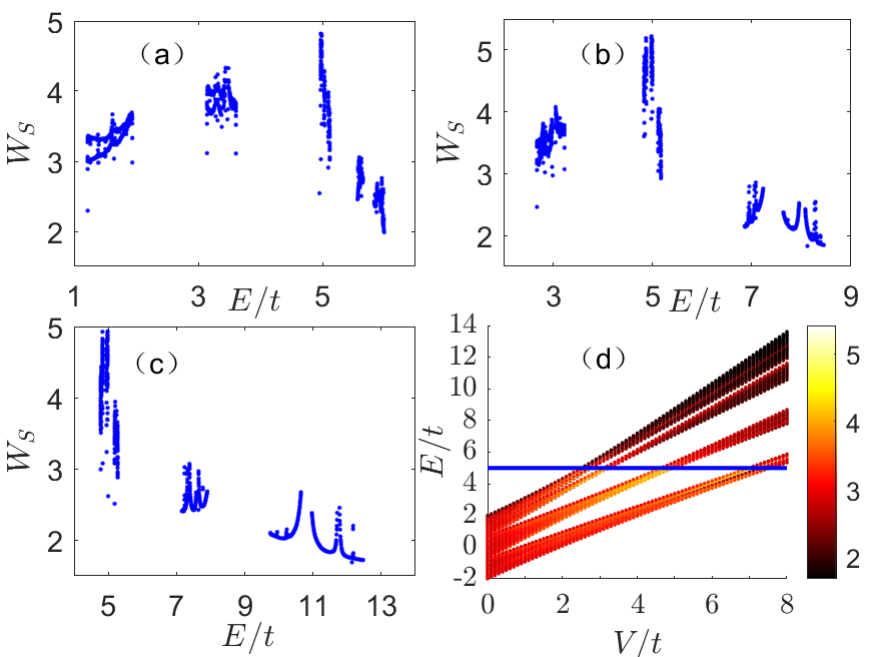}\\
  \caption{(Color Online) (a)-(c) The Wigner entropy as the change of eigenenergies. 
  (a) $V=3.2t$; (b) $V=4.8t$;  
  (c) $V=7.3t$. (d) $W_{S}$ in the $E$-$V$ parameter space with solid blue line the $E_{c}=5t$. 
  Other involved parameters are $b=0.4$ and $L=1000$.  
  }
  \label{f4}
\end{figure}

Based on the Wigner distribution, one can further obtain the Wigner entropy (marked by $W_{S}$) \cite{Wigner_entropy}. 
Considering the negativity of $W(x,p)$, the $W_{S}$ is finally calculated by the following definition
\begin{equation}
W_{S}=-\iint W(x,p) \ln{|W(x,p)|}dxdp, 
\end{equation}
where the integral region of $x$ is $\left[-L/2,L/2\right]$, and that of $p$ is $\left[-\pi,\pi\right]$.
$W_{S}$ can reflect whether the distribution of quantum states in phase space is concentrated or not. 
If Wigner entropy is small, the distribution of quantum states is concentrated. If $W_{S}$ is large, 
it means that the distribution of quantum states is relatively dispersed. We notice that the $W(x,p)$ 
of the localized and extended states is more concentrated than that of the critical states (As seen from Fig.~\ref{f3}, 
the $W(x,p)$ corresponding to the localized state and the extended state are relatively finite and concentrated on the position 
and momentum branches, respectively, while the distributions of critical sates are relatively dispersed.), 
so we infer that the quantum state has the largest $W_{S}$ at the mobility edge. In addition, we can see from Figs.~\ref{f3}(a) 
and \ref{f3}(b) that the fringe number of $W(x,p)$ of the localized state is obviously smaller than that 
of the extended state, therefore, we infer that the localized state has the smallest $W_{S}$. According to our 
inference, It seems possible to locate the mobility edge by $W_{S}$.  

To verify our inference, we plotted $W_{S}$ for $V=3.2t$, $V=4.8t$, and $V=7.3t$ in Figs.~\ref{f4}(a), \ref{f4}(b), 
and \ref{f4}(c), respectively. From the figures, we can see that $W_{S}$ peaks at $E_{c}=5t$ which means that the 
critical state has a maximal $W_{S}$ at $E_{c}$. Moreover, the $W_{S}$ below $E_{c}$ is slightly larger than that above $E_{c}$. 
Concretely, the maximal $W_{S}$ for the $V=4.8t$ case is $W^{critical}_{S}=5.2217$. Besides, 
we make averages on $W_{S}$ before and after $E_{c}$, respectively, and obtain the averaged 
$W_{S}$ for the extended states satisfies $W^{extended}_{S}=3.7875$ and that for the localized states is $W^{localized}_{S}=2.6249$.
From the numerical results, it is intuitive that $W^{critical}_{S}>W^{extended}_{S}>W^{localized}_{S}$. 
Therefore, the inference made from the distributions of $W(x,p)$ is correct under the validations of 
numerical calculation. We emphasize that this conclusion is universal to the results for other $V$ and $b$. 
In addition, the results tell that we can locate the mobility edge by using the property that the mobility edge 
produces the maximum Wigner entropy, and distinguish the extended state from the local state by comparing 
the average entropy of the left and right sides of the maximum. Meanwhile, a full phase diagram that 
contains $W_{S}$ in the $E$-$V$ parameter space has been presented in Fig.~\ref{f4}(d). The color bar
represents $W_{S}$, and from the color point of view, the mobility edge can separate the two states of extension 
and localization. 

However, the use of Wigner entropy to locate mobility edges is not an isolated case; for example, it can also be applied to the Biddle-Das Sarma model \cite{SPME_9}. This model represents a generalized AA system with long-range hopping, described by the Hamiltonian
\begin{equation}
	\hat{H}=\sum_{n,n'\neq n}te^{-p|n-n'|}    \left(\hat{c}^{\dag}_{n+n'}\hat{c}_{n}+\hat{c}^{\dag}_{n}\hat{c}_{n+n'}\right)+V_{n}\hat{c}^{\dag}_{n}\hat{c}_{n},
	\label{eq5}
\end{equation}
where $p>0$, the on-site potential is 
$V_{n}=V\cos(2\pi\alpha n)$, with $\alpha$ an irrational number, and 
$t$ is the hopping strength (set as the energy unit).
\begin{figure}[htp]
	\centering
	\includegraphics[width=0.5\textwidth]{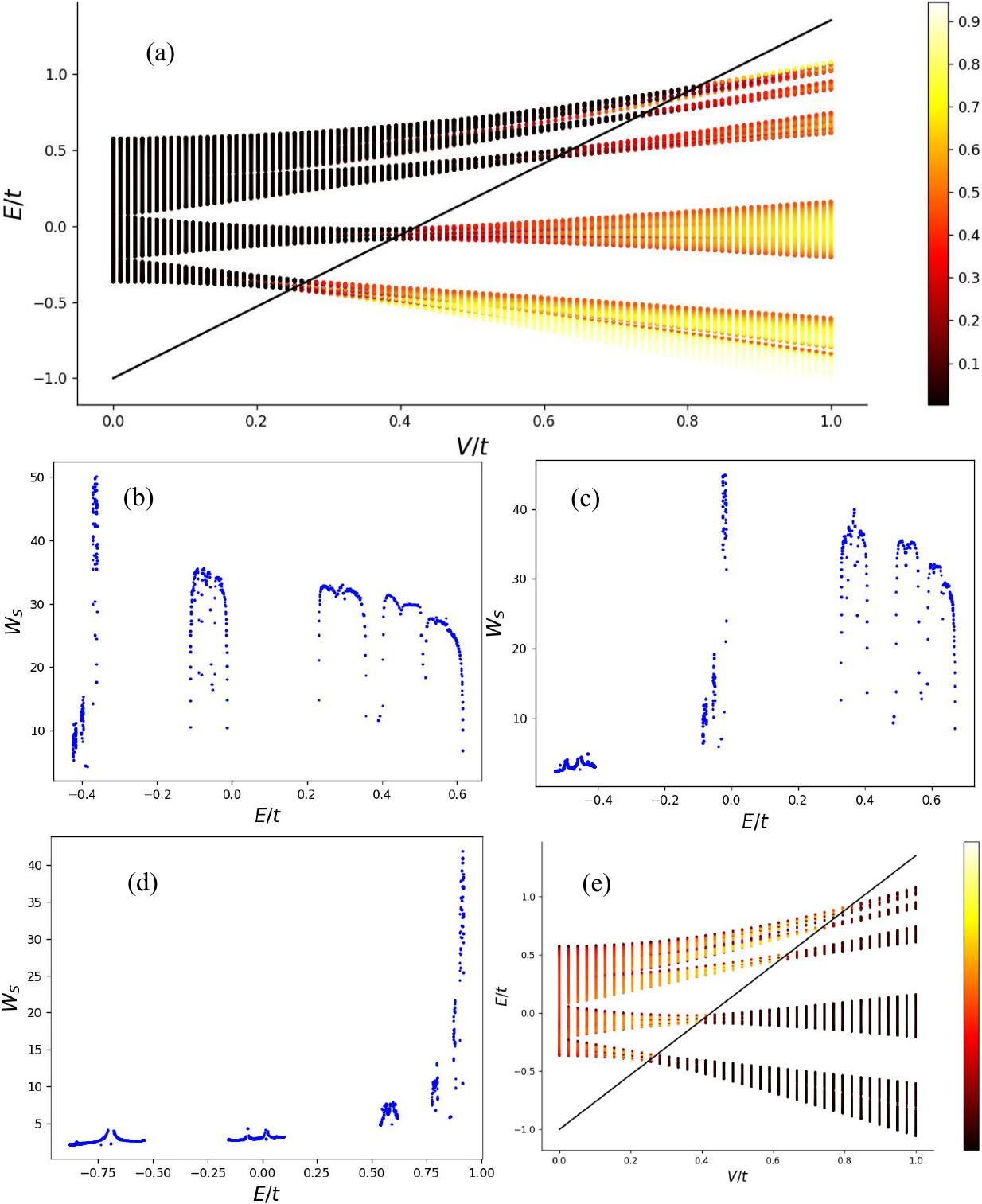}\\
	\caption{(Color Online)   (a)  Energy spectrum $E$ as a function of $V/t$, with the color bar indicating the value of the IPR; (b)-(d) Wigner entropy $W_{S}$ as a function of eigenenergy $E$ for (b) $V=0.266t$, (c) $V=0.413t$, and (d) $V=0.811t$. (e) $W_{S}$ in the $E$-$V$ parameter space, with the solid black line denoting the mobility edge $E_{c}=V\cosh(p)-t$. Other parameters are $p=1.5$ and $L=500$.
}
	\label{f45}
\end{figure}
Figure~\ref{f45} presents the energy spectrum $E$ of the Biddle-Sarma model as a function of $V/t$ for $p=1.5$ and $L=500$, with the color bar indicating the IPR. The blue solid line marks the invariant mobility edge $E_c=V \cosh (p)-t$ \cite{SPME_9}. The Wigner entropies $W_S$ are depicted for  $V=0.266t$, $V=0.413t$, and $V=0.811t$ in Figs. ~\ref{f45}(b), \ref{f45}(c), and \ref{f45}(d), respectively. In each instance, $W_S$ reaches a peak at the mobility edge $E_c$, indicating that the critical state exhibits a maximum $W_S$ value there. Moreover, the plots demonstrate that the average Wigner entropy adheres to $W^{critical}_{S}>W^{extended}_{S}>W^{localized}_{S}$. These numerical observations affirm that the Wigner distribution and entropy can effectively discriminate among extended, critical, and localized states, and that the entropy maximum serves as a robust indicator for pinpointing the mobility edge.
Recently, there are different ways to realize the measurement the Wigner distribution function, 
such as the qubit quantum processor \cite{Tilma_1,Tilma_2}, discrete atomic system \cite{Zhangtiancai}, 
photonic system \cite{Tilma_3,Songjun}, and optical parametric amplification \cite{Kalash}. 
Wigner entropy is a statistical result of Wigner distribution. With these experimental techniques, it may be possible 
to experimentally confirm our theoretical predictions in the near future.

\section{Quantum transport property}\label{S4}
In the previous study, it was proposed that the mobility edge can observed by the dynamical observation \cite{Xuzhihao} in 
an one-dimensional incommensurate system with generalized quasidisordered potential, where 
extended-critical transition can be characterized by the transition from the ballistic transport to the 
absence of diffusion. In fact, in addition to the ballistic transport and absence of diffusion, there are anomalous transport 
phenomena in the incommensurate system \cite{transport_1,transport_2,transport_3,transport_4}, such as the superdiffusive 
and subdiffusive transports. Therefore, in this section, we are concerned about whether anomalous transport occurs in the system 
with the generalized quasidisordered potential and with a mobility edge. 
\begin{figure}[htbp]
	\centering
	\includegraphics[width=0.5\textwidth]{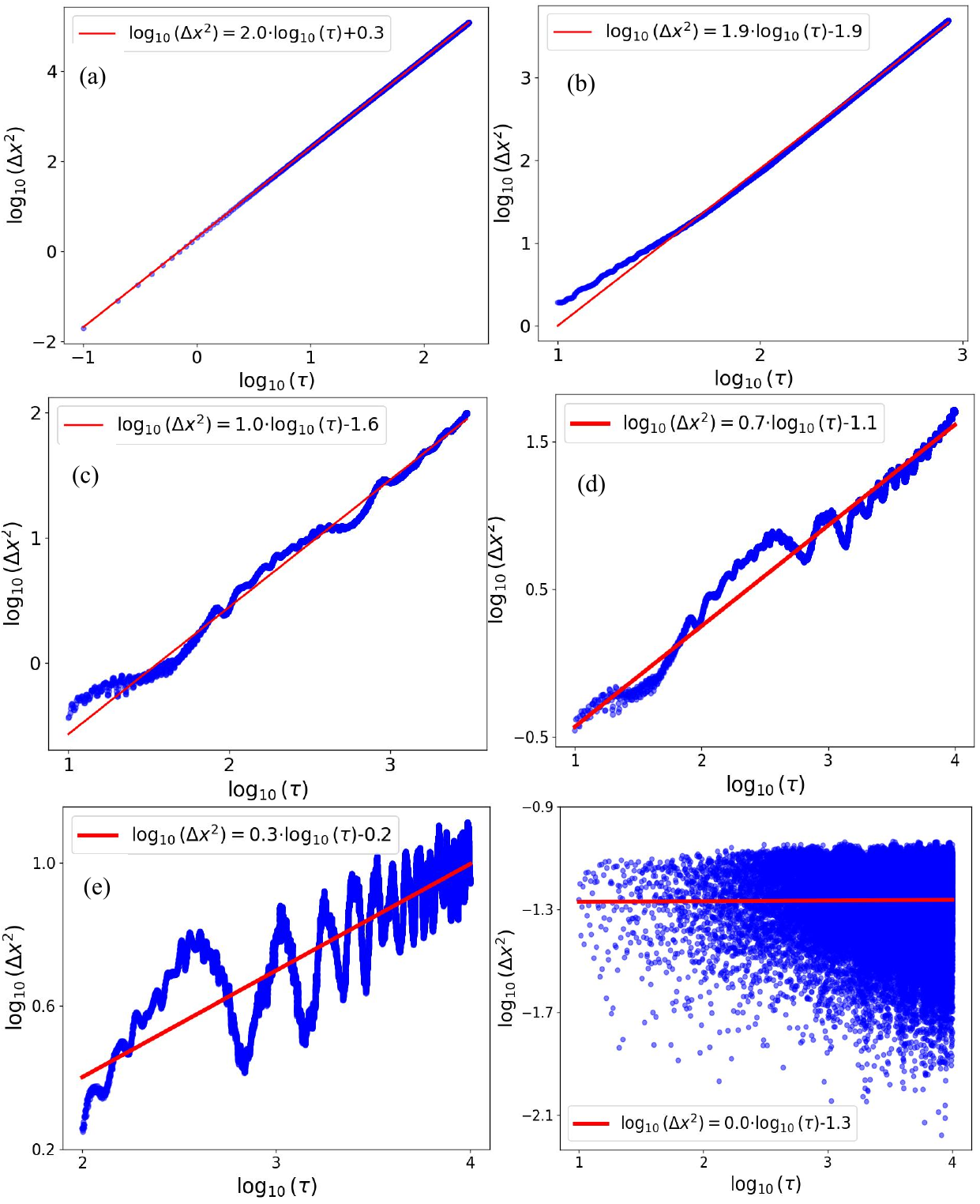}\\
	\caption{(Color Online) Time evolution of mean squared displacement $\Delta x^{2}(\tau)$ from   the system of~(\ref{eq0}) for $b=0.4$ and $L=1000$. (a) $V=0.1t$; (b) $V=5.2t$; (c) $V=7.3t$; (d) $V=7.5t$; (e) $V=7.6t$; 
		and (f) $V=12t$.  
	}
	\label{f5}
\end{figure}
\begin{figure}[htbp]
	\centering
	\includegraphics[width=0.5\textwidth]{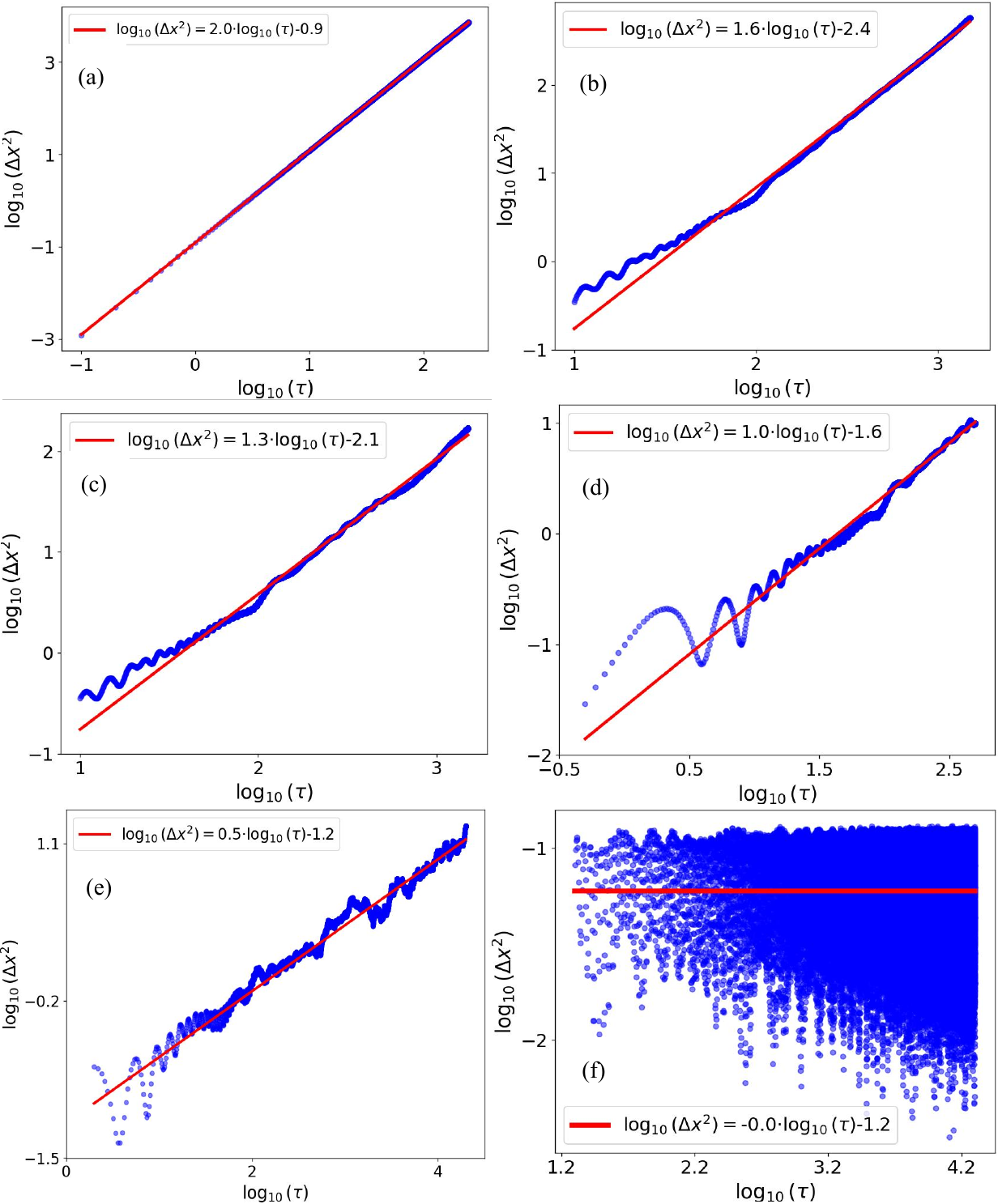}\\
	\caption{(Color Online) Time evolution of mean squared displacement $\Delta x^{2}(\tau)$ from the system of~(\ref{eq5}) for $p=1.5$ and $L=500$. (a) $V=0.05t$; (b) $V=0.65t$; (c) $V=0.7t$; (d) $V=0.75t$; (e) $V=0.85t$; and (f) $V=1.2t$.  }
	\label{f52}
\end{figure}

We analyze the transport behaviors by studying the dynamical evolution of the wave packet. The isolated generalized AA model 
is initialized with a particle occupying on site $L/2$. Thus, the initial state $\ket{\psi(\tau=0)}$ is described by the wave function 
$\ket{\psi(\tau=0)}=\sum_{n}\phi_{n}(\tau=0)\hat{c}^{\dag}_{n}\ket{0}$ where $\phi_{n}(\tau=0)=\delta_{n,L/2}$ is the Kronecker 
$\delta$ function. By solving the time-dependent Schr\"{o}dinger equation, we obtain the evolution of the wave function 
probability amplitude with time $\phi_{n}(\tau)$, and then the mean squared displacement of wave function is given by \cite{transport_2}
\begin{equation}
\Delta x^{2}(\tau)=\sum^{L}_{n=1}\left(n-L/2\right)^{2}|\phi_{n}(\tau)|^2. 
\end{equation}
The asymptotic relation of square mean displacement can also be written as 
\begin{equation}
\Delta x^{2}(\tau)\sim \tau^{\beta}, 
\end{equation}
where $\beta=2$ represents the ballistic transport, $\beta=0$ implies absence of diffusion, 
$\beta=1$ denotes diffusion, $0<\beta<1$ shows subdiffusive transport, and $1<\beta<2$ 
corresponds to superdiffusive transport. 

We first examine the wave packet dynamics in the system described by Eq.~\ref{eq0}. For parameters $b=0.4$ and $L=1000$, the mean squared displacement $\Delta x^{2}(\tau)$ is plotted for various $V/t$ in Figs.~\ref{f5}(a)-\ref{f5}(f), where the red lines denote the fitting curves using $\log_{10}(\Delta x^{2})=\beta\log_{10}(\tau)+\gamma$. The fitting parameter and associated errors are provided in Appendix \ref{app:fit_params}. As shown in Fig.~\ref{f5}(a), ballistic transport dominates for small potential strengths. With increasing $V$, the dynamics transition progressively through superdiffusive, diffusive, and subdiffusive regimes [Figs.~\ref{f5}(b)-(e)], culminating in non-diffusive behavior characterized by sharp fluctuations around the mean  [Fig.~\ref{f5}(f)] for sufficiently large  $V$.

Similar results are obtained for the Biddle-Sarma model. For $p=1.5$ and $L=500$, $\Delta x^{2}(\tau)$ is shown for different $V$ in Figs.~\ref{f52}(a)-\ref{f52}(e). Ballistic transport prevails at small 
$V$ [Fig.\ref{f52}(a)], evolving into superdiffusive, diffusive, and subdiffusive transport as $V$ increases [Figs.\ref{f52}(b)--(e)]. At large $V$, $\Delta x^{2}(\tau)$ oscillates near equilibrium, indicative of non-diffusive dynamics [Fig.\ref{f52}(f)]. These findings demonstrate that, in generalized AA models with mobility edges, transport encompasses not only ballistic, diffusive, and non-diffusive regimes but also anomalous superdiffusive and subdiffusive behaviors.

\section{Summary}\label{S5}
In conclusion, we have studied a generalized Aubry-Andr\'{e} model with tunable quasidisordered potentials. 
The model exists an invariable mobility edge that can separate the extended state from the localized state. At 
the invariable mobility edge, the wave functions are critical. We have proved that one can employ the physical image 
in phase space, i.e., the Wigner distribution, to distinguish the extended states, critical states, and the localized states. 
Based on the Wigner distribution, we can further obtain the Wigner entropy. Our numerical results show that it is also 
feasible to use Wigner entropy to distinguish extended, critical and local states. The critical state has the largest entropy, 
the extended state the middle, and the local state the smallest. Using this property, we can locate the invariable mobility edge. Meanwhile, 
this method is again verified in Biddl-Sarma model with exact mobility edge.
In addition, by studying the wave packet dynamics of the generalized AA model and the Biddle-Sarma model, we find that there are not only ballistic transport, diffusion, and 
absence of diffusion, but also are superdiffusion and subduffusion. These results further expand the extended-localized transition 
and dynamic properties of generalized quasidisordered systems. In the near future, we hope that these approaches can be 
generalized to other disordered or quasidisordered systems to distinguish between 
different forms of phase transitions.

\begin{acknowledgments}
We thank Tan Hailin for insightful discussions. This work is supported by the Natural Science Foundation of Zhejiang Province (Grants No. LQN25A040012) and 
the start-up fund from Xingzhi college, Zhejiang Normal University, and the National Natural Science Foundation of 
China (Grants No. 12174346), 
\end{acknowledgments}

\appendix
\section{Fitting Parameters}
\label{app:fit_params}

Table \ref{tab:fit_coefficients} lists the fitted coefficients 
$\beta$ and $\gamma$, along with their standard errors, for the function $\log_{10}[\Delta x^{2}(\tau))]=\beta\log_{10}(\tau)+\gamma$ , where $\beta$ and $\gamma$.

\begin{table}[htbp]
	\centering
	\caption{Coefficients and standard errors of the fitting function}
	\label{tab:fit_coefficients}
	\begin{tabular}{|c|c|c|}
		\hline
		{Fig} & {$\beta/\gamma$ Coefficient Value} & {$\beta/\gamma$ Standard Error}  \\
		\hline
		Fig6(a)    &1.9878/0.3080        & 0.0001/0.0001  \\
		\hline
		Fig6(b)    &1.9004/-1.9030       & 0.0009/0.0023  \\
		\hline
		Fig6(c)    &1.0174/-1.5884       & 0.0006/0.0020  \\
		\hline
		Fig6(d)    &0.6820/-1.1141       & 0.0008/0.0027  \\
		\hline
		Fig6(e)    &0.2982/-0.1953       & 0.0010/0.0037  \\
		\hline
		Fig6(f)    &0.0029/-1.2738       & 0.0015/0.0054  \\
		\hline
		Fig7(a)    &1.9910/-0.9143       & 0.0000/0.0001  \\
		\hline
		Fig7(b)    &1.5975/-2.3576       & 0.0008/0.0023  \\
		\hline
		Fig7(c)    &1.3402/-2.0962       & 0.0009/0.0026  \\
		\hline
		Fig7(d)    &0.9540/-1.5658       & 0.0019/0.0045  \\
		\hline
		Fig7(e)    &0.5455/-1.2127       & 0.0004/0.0015  \\
		\hline
		Fig7(f)    &-0.0000/-1.2232     & 0.0018/0.0070  \\
		\hline
	\end{tabular}
\end{table}

\bibliography{ref}

\end{document}